# Superconductivity at ~70 K in Tin Hydride SnH$_x$ under High Pressure

F. Hong[1, †, *], P.F. Shan[1, 2, †], L.X. Yang[3, †], B.B. Yue[3], P.T. Yang[1], Z.Y. Liu[1], J.P. Sun[1], J.H. Dai[1], H. Yu[1, 2], Y.Y. Yin[1], X.H. Yu[1, 2, *], J.G. Cheng[1, 2, *], and Z.X. Zhao[1, 2]

[1]*Beijing National Laboratory for Condensed Matter Physics and Institute of Physics, Chinese Academy of Sciences, Beijing 100190, China*

[2]*School of Physical Sciences, University of Chinese Academy of Sciences, Beijing 100190, China*

[3]*Center for High Pressure Science & Technology Advanced Research, Beijing, 100094, China*

[†]F. Hong, P.F. Shan and L.X. Yang have equal contribution.

* Email: hongfang@iphy.ac.cn; yuxh@iphy.ac.cn; jgcheng@iphy.ac.cn

(Dated: January 7, 2021)

## Abstract

Various tin hydrides SnH$_x$ (x = 4, 8, 12, 14) have been theoretically predicted to be stable at high pressures and to show high-critical-temperature superconductivity with $T_c$ ranging from about 70 to 100 K. However, experimental verifications for any of these phases are still lacking to date. Here, we report on the in-situ synthesis, electrical resistance, and synchrotron x-ray diffraction measurements of SnH$_x$ at ∼ 200 GPa. The main phase of the obtained sample can be indexed with the monoclinic *C2/m* SnH$_{12}$ via comparison with the theoretical structural modes. A sudden drop of resistance and the systematic downward shift under external magnetic fields signals the occurrence of superconductivity in SnH$_x$ at $T_c ≈ 70$ K with an upper critical field $\mu_0 H_{c2}(0) ≈ 11.2$ T, which is relatively low in comparison with other reported high-$T_c$ superhydrides. Various characteristic superconducting parameters are estimated based on the BCS theory.

Recently, several important breakthroughs have been achieved in fulfilling the goal of room-temperature superconductivity by compressing the hydrogen-rich materials to megabar pressures [1–5]. In 2015, Drozdov *et al.* first discovered the superconductivity in $H_3S$ with $T_c$ = 203 K at ∼ 155 GPa [1], which surpasses the record of high-$T_c$ cuprates ($T_c^{max}$ = 134 K at ambient pressure and 164 K at 30 GPa) and accelerates the explorations of hydrogen-dominated BCS superconductors. Then, the lanthanum superhydride, $LaH_{10}$, with a sodalite-like clathrate structure was successfully synthesized and the record of $T_c$ was elevated to ∼ 250-260 K at 170-190 GPa [2, 3, 6]. In this case, the dominant contribution of H electrons to the electronic density of states at the Fermi level and the strong electron-phonon coupling are responsible for the observed high-$T_c$ superconductivity [7, 8]. Subsequently, several other rare-earth superhydrides with similar structures have been synthesized [9–14] and high-$T_c$ superconductivity was discovered, *e.g.* in $YH_9$ ($T_c$ = 243 K at 201 GPa) [15] and $ThH_{10}$ ($T_c$ = 161 K at 175 GPa) [4]. Very recently, Snider *et al.* reported the observation of room-temperature superconductivity with the maximum $T_c$ = 287 K at 267 GPa in a photochemically transformed carbonaceous sulfur hydride system [5]. But the chemical composition and the crystal structure for this C-S-H system remains elusive at present.

Starting from the seminal work of Ashcroft [16], theoretical investigations based on the BCS theory played a leading role in directing the discovery of these hydride high-$T_c$ superconductors mentioned above [7, 8, 17]. As a matter of fact, nearly all binary metal hydrides have been theoretically investigated in terms of structural stability and superconducting transition temperatures [18]. Due to the technical difficulties at megabar pressures, however, only a handful of superhydrides have been experimentally investigated so far. Currently, the research interest is concentrated on the rare-earth and alkali-earth superhydrides with the clathrate structure [4, 8–10, 12, 15, 17], leaving many other binary hydrides unexplored in experiments. In this work, we turn our attention to the theoretical predictions [19–21] of novel tin hydrides $SnH_x$ having layered structures intercalated with $H_2$, $H_3$, and/or $H_4$ semi-molecular units as illustrated in Fig.1. According to the theoretical works, there are at least four thermodynamically stable phases of $SnH_x$ (x = 4, 8, 12, 14) at high pressures [19–21]. For $SnH_4$, the *Ama*2 phase (Fig. 1a) is stable in the pressure range 96-180 GPa, above which it transforms to a hexagonal $P6_3/mmc$ phase (Fig. 1b) [19]. Both phases are predicted to be superconductors with $T_c$ values of 15-22 K at 120 GPa and 52-62 K at 200 GPa for *Ama*2 and $P6_3/mmc$ phase, respectively [19]. By using evolutionary algorithm USPEX package, Esfahani *et al.* found that the hydrogen content in $SnH_x$ could be much higher at pressure above 200 GPa [21]. The *I4m*2 $SnH_8$, $C2/m$ $SnH_{12}$ and $SnH_{14}$ in Fig. 1(c-d) were found in calculations to be stable at pressures above 220, 250, and 280 GPa, respectively [21]. As shown in Fig. 1, the crystal structure symmetry of $SnH_x$ is lowered with increasing pressure, and the $H_3$ and $H_4$ molecular units appear in the lattice. For these latter phases, the predicted $T_c$ values are 81 K ($SnH_8$ at 220 GPa), 93 K ($SnH_{12}$ at 250 GPa) and 97 K (SnH$_{14}$ at 300 GPa), respectively [21]. Different from the clathrate hydrides, the electronic density of sates near Fermi level have significant contributions from both Sn and H atoms, and the intermediate frequency H-H wagging and bending vibrations make a dominate contribution to the electron phonon coupling [21].

Although the predicted $T_c$ values of SnH$_x$ are much lower than those of clathrate-type hydrides, they represent another class of binary hydrides that can have much higher H content (*e.g.* MgH$_{16}$ [22]) by incorporating a large amount of H$_2$, H$_3$, and/or H$_4$ molecular units in the lattice. It was recently proposed that electron-doping in such binary hydrides can be an effective route to achieving room-temperature superconductivity by breaking up the hydrogen molecules [7]. However, the synthesis of ternary hydrides is a challenging task. As a first step, it is necessary to investigate experimentally the stability range and superconducting properties of parent binary phases. In this work, we thus attempted to synthesize the binary SnH$_x$ by lasering heating Sn and ammonia borane (AB) at ∼200 GPa and 1700 K in a diamond anvil cell (DAC) with culet size of 50 $\mu$m. By comparing the synchrotron x-ray diffraction (XRD) pattern with the theoretically predicted structural models, the main phase of the obtained sample should be the *C*2/*m* SnH$_{12}$. In-situ resistance measurements confirm the occurrence of superconductivity at $T_c$ ∼ 70 K, in general agreement with the theoretical predictions [21]. The successful synthesis of SnH$_{12}$ in this work makes it a promising parent compound for further electron-doping investigations.

For the in-situ synthesis and the subsequent four-probe resistance measurement of SnH$_x$ at about 200 GPa, we followed the similar procedures developed in our recent work on LaH$_{10}$ [6]. A thin Sn flake of 2 $\mu$m × 10 $\mu$m × 50 $\mu$m was loaded into a DAC (culet size of 50 $\mu$m) filled with AB. After closing the DAC and applying pressure directly to ∼ 200 GPa, we heated the Sn + AB in the sample chamber to near 1700 K by using a continuous 1064-nm YAG laser from the AB side. The laser spot with an approximate diameter of 10 $\mu$m was moved forward and backward along the Sn flake between the two voltage leads, ensuring a sufficient reaction between Sn and the hydrogen released from AB. As shown in Fig. 2(a, b), the sample after laser heating at 200 GPa and 1700 K retains almost the initial configuration. The temperature was determined from the black-body irradiation spectra from the sample. The pressure values before and after laser heating were determined from the Raman signal of diamond [23, 24]. During the synchrotron XRD measurements after laser heating, we also calibrated the pressure by measuring the lattice constant of the Pt electrode near the sample. From the Raman spectra of diamond, the pressure at the culet center was determined to be about 200 GPa and it remains nearly constant after laser heating. The pressure value is also well consistent with that marked by Pt, *i.e.* ∼ 206 GPa. Synchrotron XRD at room temperature was performed on the sample in DAC at Shanghai Synchrotron Radiation Facility (SSRF, Beamline 15U) with a wavelength $\lambda$ = 0.6199 Å and beam size of ∼ 2 $\mu$m × 2 $\mu$m. The temperature-dependent resistance of the SnH$_x$ sample after XRD measurements was measured in a sample-in-vapor He$^4$ cryostat equipped with a 9 T helium-free superconducting magnet.

As mentioned above, there are at least four thermodynamically stable phases in SnH$_x$. To examine the possible phase of SnH$_x$, we performed synchrotron XRD on the sample for laser heating at ~200 GPa and 1700 K. As illustrated in the top panel of Fig. 2 (c), the collected patterns are a little bit spotty for the measurements on a small sample region, *i.e.* 2 $\mu$m × 2 $\mu$m, which indicates that the crystal size of the SnH$_x$ sample is on the scale of micron or submicron. As seen in the lower panel of Fig. 2(c), only five diffraction peaks can be well identified and there is a strong diffusing peak near $2\theta$ ~ 18 °. By comparing with the theoretically predicted structure models shown in Fig. 1, we find that the obtained SnH$_x$ sample should be dominated by *C*2/*m* SnH$_{12}$ (estimated lattice parameters: $a$ = 5.180 Å, $b$ = 3.034 Å, $c$ = 7.341 Å, $\beta$ = 149.05 °) with some extra unreacted BCC (*Im*3*m*) Sn (estimated lattice parameter $a$ = 3.036 Å). The *Ama*2 SnH$_4$ can be excluded based on previous theoretical calculations, since it can only be stable up to 180 GPa and then transforms to *P*6$_3$/*mmc* phase. The transition pressure will be much lower if considering the zero-energy effect. The Bragg position of *P*6$_3$/*mmc* SnH$_4$ provided in Fig. 2(c) corresponds to the lattice parameters at 200 GPa, given in Ref. [19]. As can be seen, the *P*6$_3$/*mmc* phase of SnH$_4$ can also be excluded given the poor matching between experimental data and the expected peak positions. In addition, the observed XRD peaks cannot be assigned to SnH$_8$ or SnH$_{14}$ [21] neither. Due to the limit number of diffraction peaks, we cannot perform structure refinement based on the *C*2/*m* SnH$_{12}$ structural model and further XRD experiment with larger sample or better crystal statistics is required to get the accurate lattice parameters. It is noted that the stability pressure range for the SnH$_{12}$ phase is lower than the prediction based on the harmonic approximation, which could overestimate the critical pressure for these phases [25, 26]. Such an "overestimation" effect was also found in LaH$_{10}$; the experimental pressure to obtain LaH$_{10}$ of ~150 GPa [2] is about 60 GPa lower than that (210 GPa) given by theoretical calculation [8].

After synchrotron XRD measurements, the same SnH$_{12}$ sample in DAC was then characterized by the resistance measurements below room temperature. Fig. 3 shows the temperature dependence of resistance *R*(*T*) during the cooling and warming processes at a slow sweep rate of about 0.3 K/min. A metallic behavior was observed at high temperatures and the resistance undergoes a sudden drop starting at ~70 K upon cooling, implying the possible occurrence of superconductivity. The resistance decreases continuously below the transition but fails to reach zero at the lowest temperatures. This is presumably due to the insufficient reaction of Sn with hydrogen in the present case employing AB rather than the pure hydrogen. During the warming-up process, the sudden jump of resistance is reproduced and shifts to ~74 K, confirming that it is an intrinsic property of the SnH$_x$ sample. As noted previously [6], the observed thermal hysteresis between the cooling and warming processes is an artifact originating from the different location of temperature sensor with respect to the sample. Because the temperature sensor was attached to the stainless-steel frame of DAC, the measured temperature is always ahead of the actual sample temperature. Thus, the real $T_c$ should be located in between 70 and 74 K. In this case, we defined the $T_c \sim 72$ K as the average of 70 and 74 K. In addition to the sharp transition, we also reproducibly observed a second small step-like anomaly in $R(T)$ slightly below $T_c$. As discussed below, this second transition might arise from part of the sample with different hydrogen content.

To confirm the occurrence of superconductivity in SnH$_x$, we performed $R(T)$ measurements under different external magnetic fields. As seen clearly in Fig. 4(a), the resistance transition is shifted gradually to lower temperature upon applying magnetic fields, which is a hallmark of the superconducting transition. In addition, the superconducting transition is broadened up and the step-like anomalies become invisible with increasing fields. It is noted that the superconducting transition temperature decreases by 2-3 K at zero field in comparison with the data shown in Fig. 3. This should be ascribed to the pressure change after the first thermal cycling. Indeed, we confirmed that the pressure drops to ~190 GPa after the first resistance measurements shown in Fig. 3. But the pressure in DAC does not change any more after further thermal cycling as checked after all the measurements in magnetic fields were finished.

As seen in Fig. 4(a), the effect of magnetic field on the superconducting transition is very strong; it is lowered by ~35 K under 7 T. To quantify the influence of magnetic field on $T_c$, here we determined the $T_c$ value as the middle-point temperature of the resistance drop and plotted in Fig. 4(b) as a function of external fields. First, we can estimate the zero-temperature upper critical field $\mu_0H_{c2}(0)$ by fitting the $\mu_0H_{c2}(T)$ with the empirical Ginzburg–Landau (G-L) equation, viz. $\mu_0H_{c2}(T) = \mu_0H_{c2}(0)(1 - t^2)/(1 + t^2)$, where $t$ is the reduced temperature $T/T_c$. As shown by the solid green curve in Fig. 4(b), the G-L fitting yields the $\mu_0H_{c2}(0) = 11.2$ T and $T_c = 67.3$ K. The obtained $\mu_0H_{c2}(0)$ is much smaller than the Pauli paramagnetic limited $\mu_0H_p(0) = 1.84\,T_c \approx 122$ T, implying that the spin-paramagnetic effect is the major pair breaking mechanism. Based on the obtained $\mu_0H_{c2}(0)$, we can estimate the G-L coherent length $\xi = 5.4$ nm according to the formula $\xi = (\Phi_0/2\pi H_{c2}(0))^{1/2}$, where $\Phi_0 = 2.068 \times 10^{-15}$ Wb is the magnetic flux quantum. By assuming that SnH$_x$ is a weak type II superconductor, the Fermi velocity $v_F = 2.6 \times 10^5$ m/s can be estimated by using the formula: $v_F = \pi\xi(0)\Delta(0)/\hbar$, where $\Delta(0) = 1.76\,\kappa_BT_c \approx 10.2$ meV is the superconducting gap energy in the weak coupling BSC limit. The obtained $v_F$ is very close to that of LaH$_{10}$. Under the free electron gas approximation, the carrier density can be calculated to be $n = 6.63 \times 10^{27}$ m$^{-3}$ according to $m^*_{eff}\,v_F = \hbar(2\pi n)^{1/3}$. Here, the $m^*_{eff}$ can be estimated from Eliashberg calculations $m^*$ $_{eff}$ = (1 + $\lambda$) $m_e$ [27], where the electron-phonon coupling constant $\lambda \sim 1.25$ [21]. The obtained carrier density of LaH$_{10}$ is about $n = 3.45 \times 10^{28}$ m$^{-3}$ [28], which is about an order of magnitude larger than SnH$_x$. The Fermi energy is estimated to be $E_F = m^*_{eff} v_F^2/2 \sim 0.43$ eV, lower than that of LaH$_{10}$, i.e. 0.96 eV [28]. From the initial slope of $H_{c2}(T)$ at $T_c$, i.e. $dH_{c2}/dT|_{Tc}$ = -0.21 T/K, we can also obtain the orbital limited $\mu_0 H_{c2}^{orb}(0)$ = -0.69 × $T_c$ × $dH_{c2}/dT|_{Tc}$ = 9.6 T in the dirty limit of Werthamer–Helfand–Hohenberg (WHH) model without spin–orbit coupling [29]. Based on these results, we can evaluate the Maki parameter $\alpha = \sqrt{2}H_{c2}^{orb}(0)/H_p(0) = 0.11$. From these above characterizations, we can reach the conclusion that SnH$_{12}$ obtained at 200 GPa and 1700 K is a weak type-II BCS superconductor with low carrier density.

In the study by Esfahani *et al.*, the predicted $T_c$ values for SnH$_x$ fall in the range between 81 and 97 K depending on the hydrogen content [21]. For *C2/m* SnH$_{12}$, its $T_c$ is predicated to be 93 K at 250 GPa in case of Coulomb pseudo-potential $\mu^* = 0.1$. In our experiment, the observed $T_c$ is about 72 K, which is ~21 K lower than the theoretical value, if we ignore the pressure effect. Similar situation also occurs in LaH$_{10}$, for which the predicted $T_c$ is 286 K by using $\mu^* = 0.1$ [8] whereas the experimental $T_c$ is 250-260 K [2, 3, 6]. For LaH$_{10}$ with the clathrate structure, the H atoms are weakly bonded covalently with H-H distance of 1.08 Å at 200 GPa and the molecular H$_2$ unit is strongly suppressed [17]. The weak H-H interaction in LaH$_{10}$ with large distance (comparable with the H-H distance in atomic metallic hydrogen at 500 GPa) is beneficial for the electron-phonon coupling (EPC) since all H vibrations are involved in the EPC process [17]. In contrast, SnH$_{12}$ contains a substantial amount of H$_2$ and H$_4$ semi-molecule units as presented in Fig. 1(d), and the latter is formed by two H$_2$-groups with a distance about 0.99 Å at 250 GPa [21]. Hence, EPC effect in SnH$_{12}$ is not as strong as that in LaH$_{10}$. In addition, SnH$_{12}$ has less H-projected density of states at the Fermi level [21]. All these factors mentioned above determine the relatively low $T_c$ of SnH$_{12}$ in comparison with the LaH$_{10}$. Nonetheless, the successful synthesis of SnH$_{12}$ and confirmation of its superconductivity in this work makes it a promising parent compound for exploring novel ternary hydrides with higher $T_c$. According to the recent proposal by Sun *et al.* [7], if we can dope electrons into such hydrides and break the H$_2$/H$_4$ molecules, it is expected to enhance the $T_c$ substantially.

In summary, we have synthesized SnH$_x$ sample by laser heating the Sn and ammonia borane inside a diamond anvil cell at about 200 GPa and 1700 K. The obtained sample could be the theoretically predicted *C2/m* SnH$_{12}$ and is confirmed to be Type-II BCS superconductor with $T_c \approx 70$ K and a relatively low upper critical field $\mu_0 H_{c2}(0) = 11.2$ T. We have also estimated some characteristic superconducting parameters and compared with those of clathrate-type LaH$_{10}$. Our results indicate that the SnH$_{12}$ could be a promising parent compound for exploring novel electron-doped ternary hydrides with higher $T_c$.

# Acknowledgments

The X-ray diffraction was done at Beamline 15U (supported by Lili Zhang and Xiaosheng Lin) of Shanghai Synchrotron Radiation Facilities. This work is supported by the Strategic Priority Research Program of the Chinese Academy of Sciences (Grant Nos. XDB33000000 and XDB25000000), the Beijing Natural Science Foundation (Grant No. Z190008), the National Natural Science Foundation of China (Grant Nos. 11575288, 12025408, 11888101, 11904391, 11834016, 11874400, 11921004, U1930401), the National Key R&D Program of China (Grant Nos. 2016YFA0401503 and 2018YFA0305700), the Youth Innovation Promotion Association, the Key Research Program of Frontier Sciences and the Interdisciplinary Innovation Team of Chinese Academy of Sciences (Grant Nos. 2016006, QYZDBSSW-SLH013, and JCTD-2019-01). Some instruments used in this study were built for the Synergic Extreme Condition User Facility.

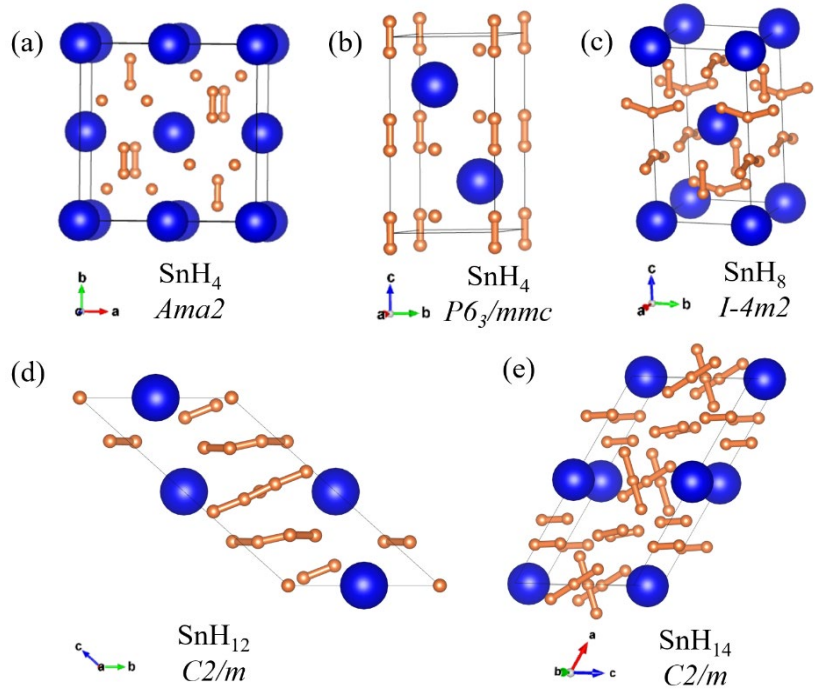

FIG. 1. (color online) Theoretical structure models proposed for various $SnH_x$. (a) $SnH_4$ with space group *Ama*2 (96-180 GPa) [19], (b) $SnH_4$, $P6_3/mmc$ (> 180 GPa) [19], (c) $SnH_8$, $I4m2$ (> 220 GPa) [21], (d) $SnH_{12}$, $C2/m$ (> 250 GPa) [21], (e) $SnH_{14}$, $C2/m$ (> 280 GPa) [21].

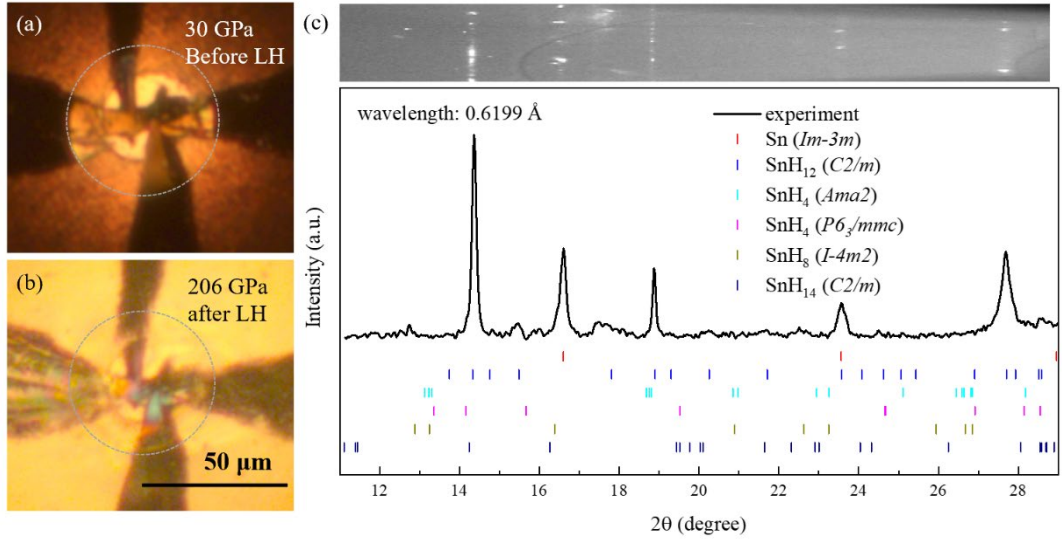

FIG. 2. (color online) The photographs of sample inside DAC and synchrotron x-ray diffraction data for SnH$_x$ sample at 206 GPa. (a, b) The optical photos of SnH$_x$ sample before and after laser heating; the dotted circles indicate the approximated edge of culet; (c) x-ray diffraction data: upper panel, the original XRD data; lower panel, integrated XRD data with the Bragg positions of various Sn hydride phases [19, 21] plotted below the experimental data for comparison. By comparing the experimental data and theoretical structure models, the XRD pattern could be assigned to $C2/m$ SnH$_{12}$ and $Im3m$ Sn.

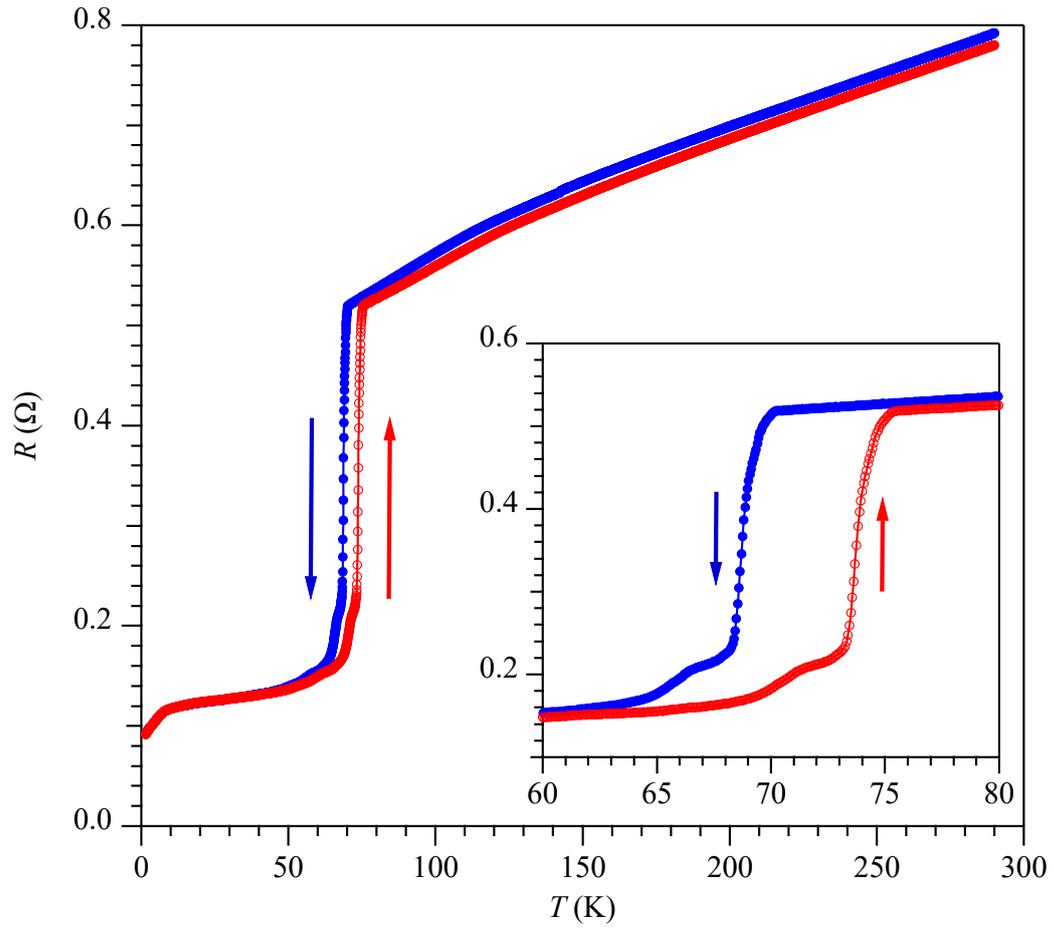

FIG. 3. (color online) *R-T* curve at ~200 GPa from 2 K to 290 K at zero field during the cooling (blue) and warming (red) processes. Inset shows the zoom-in region between 60 and 80 K.

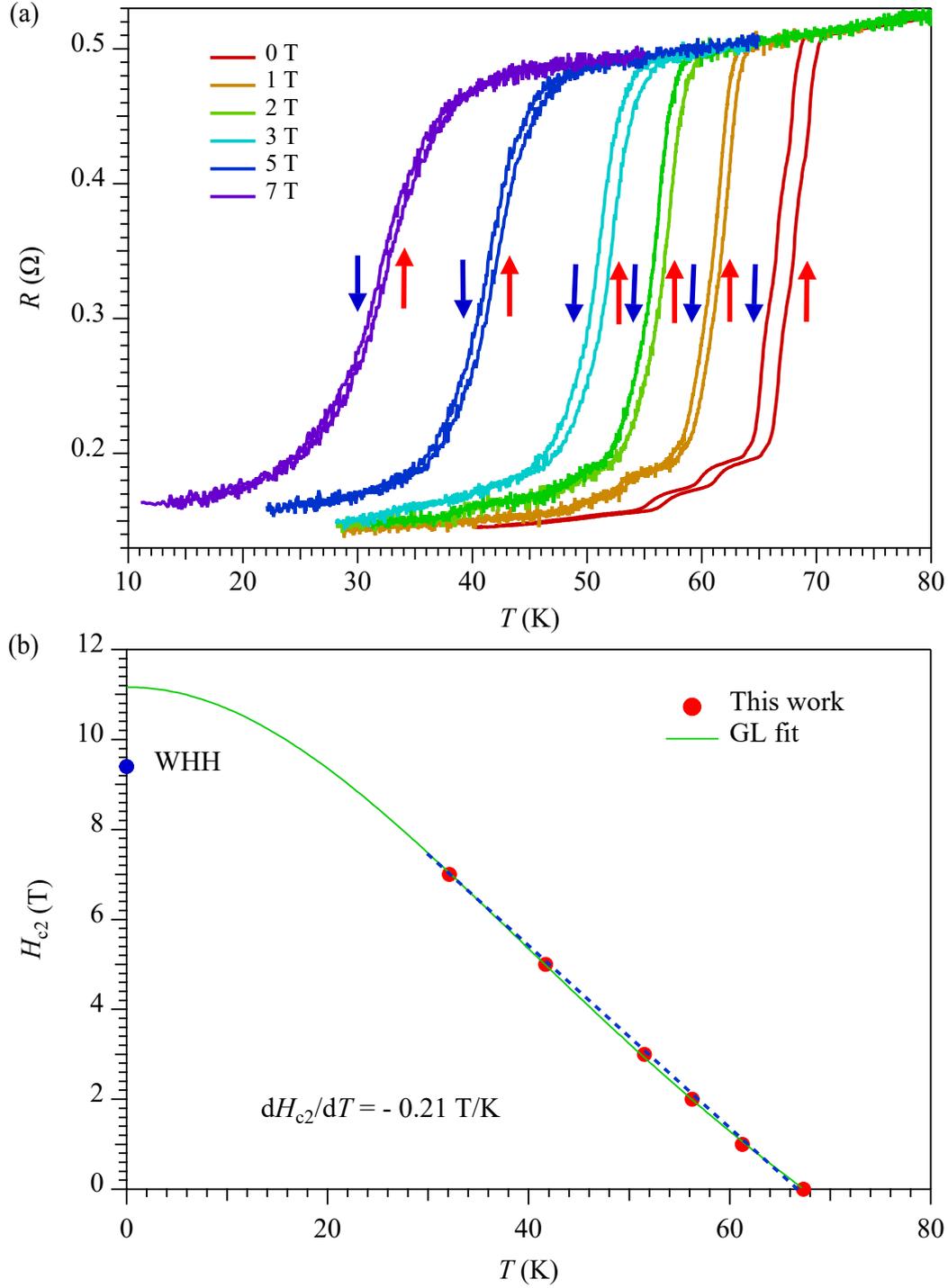

FIG. 4. (color online) Magnetic field effect on the superconducting transition of SnH$_x$ at ~190 GPa. (a) temperature dependence of resistance near the superconducting transition under different magnetic fields; (b) Field dependence of the critical temperature fitted by the Ginzberg–Landau (green solid curve) and Werthamer–Helfand–Hohenberg (WHH) (blue dashed line) expressions.

13